# Nanomechanical Characterization of an Antiferromagnetic Topological Insulator

Shuwan Liu,[1] Su Kong Chong,[2] Dongwook Kim,[3] Amit Vashist,[1] Rohit Kumar,[1] Seng Huat Lee,[4,5] Kang L. Wang,[2] Zhiqiang Mao,[4,5] Feng Liu,[3] Vikram V. Deshpande[*,1]

[1]Department of Physics and Astronomy, University of Utah, Salt Lake City, Utah 84112, United States
[2]Department of Electrical and Computer Engineering, and Department of Physics and Astronomy, University of California, Los Angeles, California 90095, United States
[3]Department of Materials Science and Engineering, University of Utah, Salt Lake City, Utah 84112, United States
[4]Department of Physics, The Pennsylvania State University, University Park, Pennsylvania 16802, USA
[5]2D Crystal Consortium, Materials Research Institute, The Pennsylvania State University, University Park, Pennsylvania 16802, USA
*Corresponding author: E-mail: vdesh@physics.utah.edu; Telephone: 801- 581-6570

**Abstract**: The antiferromagnetic topological insulator $MnBi_2Te_4$ (MBT) exhibits an ideal platform to study exotic topological phenomena and magnetic properties. The transport signatures of magnetic phase transitions in the MBT family materials have been well-studied. However, their mechanical properties and magneto-mechanical coupling have not been well-explored. We use nanoelectromechanical systems to study the intrinsic magnetism in MBT thin flakes via their magnetostrictive coupling. We investigate mechanical resonance signatures of magnetic phase transitions from antiferromagnetic (AFM) to canted antiferromagnetic (cAFM) to ferromagnetic (FM) phases versus magnetic field at different temperatures. The spin-flop transitions in MBT are revealed by frequency shifts of mechanical resonance. With temperatures going above $T_N$, the transitions disappear in the resonance frequency map, consistent with transport measurements. We use a magnetostrictive model to correlate the frequency shifts with the spin-canting states. Our work demonstrates a technique to study magnetic phase transitions, magnetization and magnetoelastic properties of the magnetic topological insulator.

## INTRODUCTION

Nanoelectromechanical systems (NEMS) have many applications in nano filtering, ultra-high sensitivity mass/force/pressure sensing, and molecular detection. Two-dimensional NEMS have been extensively studied in graphene[1–9] for high-performance devices due to their ultra-low mass density and ultra-high mechanical flexibility and stiffnesses. Besides graphene, other 2D materials have also seen applications in NEMS for their unique mechanical, electrical, and optical properties with different degrees of freedom, such as optical properties and charge density wave transitions of transition-metal dichalcogenides[10–12] and magnetic phase transitions in antiferromagnetic/ferromagnetic materials[13]. Recently, the field-induced magnetostriction[14–16] in an antiferromagnetic 2D material has been used to develop new control of magneto-mechanical coupling in the context of NEMS[17].

Another interesting layered material is the intrinsic magnetic topological insulator $MnBi_2Te_4$ (MBT)[18–20] in which spins are coupled intralayer in the ferromagnetic (FM) state and interlayer coupling is antiferromagnetic (AFM) in c axis. The antiferromagnetic topological insulator MBT material family exhibits an excellent platform for studying new electronic and magnetic couplings with mechanical motion. The introduction of magnetism in the topological material MBT leads to exotic phenomena such as the quantum anomalous Hall effect[21,22], axion insulator state[23,24] and layer Hall effect[25]. The band structures of the MBT family of compounds have been well studied using angle-resolved photoemission spectroscopy (ARPES) measurements[19,26]. These measurements revealed that MBT is a heavily electron-doped compound, and the Fermi level lies in the conduction band. This could be due to antisite defects or nonstoichiometry. Recent first principles calculations, ARPES and transport measurements have shown that Fermi level of $Mn(Bi_{1-x}Sb_x)_2Te_4$ (MBST) can be tuned from conduction band to valence band by varying the Sb concentration $x$[27,28].

While the electrical and magnetic properties of MBT have been well studied, their coupling with mechanical degrees of freedom is less studied, especially in nanoscale devices. Previous research on MBT films has shown the importance of magnon-phonon interactions and substrate interaction for magnetic phase transitions via Raman spectroscopy measurements[29–31]. Here we report the exchange magnetostriction effect in the Mn $(Bi_{1-x}Sb_x)_2Te_4$ (x = 0.2) film. The exchange magnetostriction in the antiferromagnetic insulator MBST responds to the magnetization of $Mn^{2+}$ and induces magnetostrictive strain in the film. The changes of magnetostrictive effect induced strain correspond to the changes in resonance frequency. The magnetoelastic energy is determined from the magnetization due to the external field. Thus, we use 2D NEMS to study magnetic phase transitions in the MBT material family. Furthermore, the suspended membrane geometry of nanoelectromechanical resonator devices eliminates the interaction between the sample and the substrates and may provide cleaner manifestations of these phase transitions[32].

## RESULTS AND DISCUSSION

The substrate for the nanoelectromechanical (NEM) resonator device was prepared in the following steps[33]: (*i*) the bottom gate gold electrodes (30 nm) were photolithographically-patterned on a highly resistive silicon wafer (600 µm), (*ii*) the $SiO_2$ layer (300 nm) was deposited by plasma-enhanced chemical vapor deposition, (*iii*) the source and drain gold leads (25 nm) were patterned and deposited as in (*i*), (*iv*) a circular hole was etched down to the bottom gate electrode by a dry reactive-ion etching process. (*v*) The MBST flakes were exfoliated and transferred onto the patterned gold leads over the circular hole using a micromanipulator transfer stage. First, we studied the basic mechanical properties of MBST sample. Figure 1a shows the optical image of a representative MBST NEM resonator device with a thickness about 30 nm. The MBST flake is suspended over a hole with a radius of 4 µm. The underneath circular trench is more visible for the thinner MBST flakes. The schematic cartoon in Figure 1a illustrates the side view of the device structure in which the MBST membrane is suspended over the hole. The two-terminal resistance of the MBST flake is about 5 kΩ at room temperature. The MBST flake and the underneath gold gate electrode form a parallel capacitor structure with vacuum as dielectric.

Radio frequency (RF) measurements were performed at temperature of 1.5 K. Figure 1b shows the circuit diagram of the RF measurement. Mechanical vibrations are driven by a RF signal outputting from a vector network analyzer (VNA). The RF signal is coupled to a DC gate voltage through a bias tee. The gate voltage applies an electrostatic force which causes mechanical tension in the MBST membrane. The transmitted RF signal is decoupled using a second bias tee before feeding back to the VNA. The fundamental resonance mode of the MBST membrane can be fitted to a Lorentzian line shape at a peak frequency of about 32.7 MHz, where the quality factor can reach up to 10,000 (Supplementary Information 1, Figure S1). Figure 1c shows the color plot of the RF transmission ($S_{21}$) as functions of frequency and gate voltage. The original plot without fitting curve is shown in Figure S2. We extracted the gate dependent resonance frequencies ($f_{res}$) and fitted them with a continuum mechanics model[34] for a fully clamped membrane (Supplementary Information 3). The fitting result is displayed in solid black line as depicted in Figure 1c. The resonance frequency decreases to its minimum at $V_g$ ~ 80 V and then increases with further increase of $|V_g|$, similar to other NEMS[35]. We further determined the fitting parameters for the MBST NEMS such as the Young's modulus, mass density (~7.2 kg/m$^3$), and the built-in stress (~88 N/m, Supplementary Information 8). To supplement this data, we performed DFT calculations (Supplementary Information 5, Table S2) for the basic parameters of Young's modulus and Poisson's ratio in different doping levels (Sb = 0,0.5,1) of MBST and different induced strains. We chose the fitting parameters of Young's modulus E = 84.4 GPa, and Poisson's ratio $v_P$ = 0.21 and got the built-in strain $\epsilon_0 \approx 0.83\%$ in our nanomechanical resonator device based on these parameters.

Next, we investigated the effect of a perpendicular magnetic field on the $f_{res}$ for the MBST NEMS device. Figure 1d shows the color plot of $S_{21}$ as functions of frequency and magnetic field with $V_g$ fixed at 70 V. Two transitions are observed in magnetic field. The first one is associated with a jump in the $f_{res}$ at around 3.7 T. The second one happens at around 8T beyond which the $f_{res}$ saturates with increasing magnetic field. These transitions are highly symmetric with the magnetic field. The transitions in $f_{res}$ appear in line with the magnetic phase transitions in the MBST[26,32,36]. The first transition involves a spin-flop process from AFM to cAFM phase, while the second transition corresponds to cAFM to FM transition. We have seen similar results in all (six) devices measured so far (see Supplementary Information 10 for additional data).

For comparison, we performed transport measurements on MBST flakes with similar thicknesses. Figure 2a and b show the gate and magnetic field dependence of $R_{xx}$ and $R_{xy}$ for a 30 nm MBST Hall bar device (Figure 2d, inset). By applying top gate voltage, the chemical potential of the MBST is tuned around the charge neutrality. The kinks or vertical steps in $R_{xx}$ color map at ±2.8 T are assigned to the $H_{c1}$ corresponding to the spin-flip transition from AFM to cAFM state. The $H_{c1}$ is nearly independent of the top gate voltage and thus the carrier density. The $R_{xy}$ color map presents the Hall signal at different magnetic phases as controlled by the magnetic field. The anomalous Hall signal develops in FM (partially in cAFM) phase when the top gate voltage is controlled close to charge neutrality ($V_{tg}$ > +12 V). The line cuts of $R_{xx}$ and $R_{xy}$ taken at different temperatures in the two distinct Hall signal regions at $V_{tg}$ of +7.8 V (hole carrier) and +16.5 V (charge neutrality) are plotted in Figure 2c-f. The positive slope in $R_{xy}$ at magnetic field |H| < |$H_{c1}$| for both carrier densities is attributed to hole-type bulk carriers as a result of Sb substitution[28,36,37]. The evolution of $H_{c1}$ and $H_{c2}$ with temperature for the two carrier densities are tracked by the blue and red arrows, respectively. The $H_{c1}$ is determined by the sharp kink in $R_{xx}$, whereas the $H_{c2}$ is indicated by the second bending point in the $R_{xy}$ curves. The $R_{xx}$ kink fades with the increase in temperature and eventually disappears at 25 K, which is consistent with the Néel temperature (~23 K)[36,37] of MBST.

To confirm the correlation between the $f_{res}$ changes and the magnetic transitions, we conducted magnetic field dependent RF measurements for the MBST resonator device at different temperatures as summarized in Figure 3. The disappearance of the transitions beyond the Neel temperature $T_N$ ~ 23 K agrees with the transport results. The monotonic increasing of $f_{res}$ with magnetic field at temperature of 25 K suggests the absence of long-range magnetic (spin) order as in paramagnetic (PM) phase. Also, the evolution of $H_{c1}$ and $H_{c2}$ with temperature below $T_N$ based on the change in $f_{res}$ is summarized in Figure S5. Similar to the transport data, both the $H_{c1}$ and $H_{c2}$ suppress, albeit at different rate, with the increase in temperature. At temperature 10 K, the $f_{res}$ is no longer saturating at high field, but increases linearly with magnetic field beyond the $H_{c2}$. The positive slope in $f_{res}$ follows the same pattern qualitatively as the magnetic field dependent $f_{res}$ trend at 25 K. This indicates that the magnetic field will not prolong the FM state at higher temperature, similar to the observations in magnetic force microscopy[38]. In contrast, the magnetic field dependent $f_{res}$ preserves the similar function in the AFM (as well as cAFM) phase at different temperatures, indicating the robust AFM ground state in the MBST. At higher temperature, the PM phase dominates the high magnetic field regime while the cAFM phase is nearly suppressed.

We extracted the frequency data with the magnetic field and temperature dependence from Figure 3a-e and obtained a phase diagram. Figure 3f depicts the 3D plot of a magnetic phase diagram based on normalized frequency (measured frequency at different magnetic fields divided by the frequency at zero magnetic field). The x and y axes are temperature and magnetic field respectively, with projection on the x-z plane being the low-temperature limit and the y-z plane corresponding to B = 9 T, respectively. The three magnetic phases viz. AFM, cAFM and FM state are indicated; with temperature increasing, the PM phase emerges. The magnetic field dependence and temperature dependence of resonance frequency both strongly support the magnetic phase transition picture in the MBST sample. The dependence of resonance frequency on the magnetic field indicates a magnetostrictive resonator. The mechanical resonance frequency is correlated with the internal magnetic order, which changes with the external magnetic field. Around $H_{c1}$ at base temperature, the resonant frequency has an abrupt jump from 32.75 MHz to 32.645 MHz. The frequency change is about 105 kHz, and the change rate is about 0.32% ($\frac{f_{AFM}-f_{cAFM}}{f_{AFM}} \approx 0.32\%$). In the canted AFM state, the change of resonance frequency correlates with the canting angle of the spin. We use the magnetostriction model to fit our experimental data. The model of magnetostriction indicates that the competition between minimizing the internal magnetic energy and elastic energy contributes to the resonance frequency shifts with magnetic state[17]. The elastic energy of the membrane per unit area can be expressed as $U_{el} = -\frac{1}{2}\sigma_0\epsilon$, where $\epsilon$ is the strain and $\sigma_0$ is the stress. In particular, the free energies per unit area for the free membrane in the AFM, canted AFM and FM state in the zero-temperature limit are:

AFM state:
$$F_{AF} = nF_0 + (n-1)J_\perp, \qquad (1)$$

Canted AFM state:
$$F_{cAF} = nF_0 + (n-1)J_\perp \cos\theta + \frac{1}{2}nK_{eff}\sin^2\theta, \qquad (2)$$

FM state:
$$F_{FM} = nF_0 - (n-1)J_\perp - \mu_0 M_0 \left(H_\perp - \frac{M_0}{t}\right), \qquad (3)$$

where $\theta$ is the canting angle of the spin of Mn atoms, $J_\perp$ is the interlayer exchange coupling with energy per unit area and $K_{eff}$ is the anisotropy with effective energy per unit area per layer.

The total energy of the membrane is the sum of elastic energy, the free energy of the membrane and the boundary energy of the fully clamped drum. By taking the derivative of the total energy with respect to strain and set the derivative equation to zero to minimize total energy, we get the frequency changes between AFM state, canted AFM state and FM state as a function of magnetostrictive coefficients and spin canting angle Comparing the model with the experimental data for frequency shifts, we get $\frac{\partial J_\perp}{\partial \epsilon} = 0.0094$ N/m. Using unit cell area $A_u \approx 0.16\ nm^2$ and $n = 31$ as layers corresponding to the reasonable estimate for the thickness (42 nm) of the sample, we obtain one magnetostrictive coefficient: $\frac{\partial J_\perp}{\partial \epsilon} A_u \approx 9$ meV. By comparing the fractional change of resonance frequency in canted AFM state and AFM state, we obtain the other magnetostrictive coefficient: $\frac{\partial K_{eff}}{\partial \epsilon} A_u \approx 0.6$ meV [Supplementary Information 8]. Using the values of the two magnetostrictive coefficients obtained above, we can fit the resonance frequency as a dependence of magnetization. For the resonance frequency we get from the experiment, we can extract the magnetization (absolute value) as a function of the magnetic field. It is shown in Figure 4a. The magnetization increases with the magnetic field before saturation. In a small range of magnetic field, the magnetization is small, almost zero; when the magnetic field is near $H_{c1}$, there is a dramatic increase in magnetization. The magnetization increases monotonically with the magnetic field in the canted AFM state and finally saturates in the FM state.

We have also used DFT calculations to study the stability of the phases and obtain theoretical estimates for the magnetoelastic coefficients obtained above. Figure 4b shows the magnetic phase transitions under the external magnetic field. The relative energy [$\Delta E(\theta)$] of AFM versus FM configuration is obtained as the spin is rotated from AFM to FM state. The insets show the spin directions of each data point. The effect of the external magnetic field is considered by adding an energy gain [$\mu_B (\frac{m_{1,z}+m_{2,z}(\theta)}{2})B$] (index 1 and 2 indicates the top and bottom SL, respectively). As the external magnetic field increases, the relative energy of spin configuration aligned to the external field is lowered by this energy gain, as presented by different colored lines. Above the critical field ($B_c$=3.509 T, green line), the relative energy of FM states becomes negative, indicating the FM state becomes energetically favorable over the AFM state. Red and blue shaded areas in Fig. 4b show where the AFM and FM states are more stable, respectively. For the DFT results of derivative of exchange and anisotropy with strain for MBST sample we obtain: $\frac{\partial J_\perp}{\partial \epsilon} Au \approx 2.8$ meV, $\frac{\partial K_{eff}}{\partial \epsilon} Au \approx 0.2$ meV [Supplementary Information 9]. The experimental and theoretical results for the two magnetostrictive coefficients are fairly consistent in that they are of the same order, with the interlayer exchange coupling in the few meV range and the anisotropy energy being one order of magnitude smaller.

## CONCLUSION

We have demonstrated an exchange magnetostriction effect in magnetic topological insulator MBST. We first studied the mechanical properties of MBST resonator, such as the elastic modulus, built-in strain, and Q factor. The strain tuning of resonance frequency was manipulated by the external magnetic field. The magnetostrictive behavior in the suspended MBST film yielded information on different magnetic states in the magnetic phase diagrams. We modified the magnetostriction model to quantitatively describe the spin flop transition in the canted AFM state of MBST and extracted the magnetization as a function of the magnetic field from mechanical resonance. Our results quantitatively establish the correlation between intrinsic magnetic ordering and resonance frequency in the case of MBT. Furthermore, we quantitatively obtained the exchange and anisotropy energies concerning strain. Overall, our magneto-mechanical coupling study provides a new platform to explore the inherent magnetism of MBT family materials.

## METHODS

**Crystal Growth**. Single crystals of Mn(Bi$_{1-x}$Sb$_x$)$_2$Te$_4$ were synthesized using the Te-flux method[28]. The manganese powder, bismuth shot, antimony shot, and tellurium lumps with a molar ratio of 1:5(1-x):5x:16 were loaded in an alumina crucible and sealed in a quartz tube under a high vacuum. The mixture was heated to 900°C for 12h to promote the homogeneous melting and slowly cooled down to a temperature window of 590°C to 630°C at a rate of 1.5°C/hour. The platelike single crystals can be obtained after removing the excess Bi-Te flux by centrifugation.

**DFT Calculations.** The first principles calculations are performed within the framework of density functional theory (DFT) by Vienna an-initio calculation package (VASP)[39] in Perdew-Burke-Ernzerhof-type (PBE) generalized gradient approximation (GGA) [J. P. Perdew, K. Burke, and M. Ernzerhof, Generalized Gradient Approximation Made Simple[40]. To address the electron correlation due to the Mn 3d orbitals, we include a Hubbard U-J parameter 4eV. The energy cutoff for the plane-wave basis is set to 450 eV, and K-point sampling is done using a 12×12×2 Monkhorst-Pack grid[41]. All the crystal structures are fully relaxed until the atomic forces are below $1e^{-2}$ eV/Å. vdW correction is included by the DFT-D3 method[42] to better describe the interlayer dispersion forces.

## ASSOCIATED CONTENT

The Supporting Information is available free of charge at ACS website.

## AUTHOR INFORMATION

### Author Contribution

S.W.L. performed the measurements, analyzed the data, and co-wrote the manuscript under the supervision of V.V.D.; S.K.C. helped with sample preparation and performed the transport measurements under the supervision of K.L.W.; D. K. performed DFT calculations under the supervision of F.L.; A.V. helped with measurements; R.K. fabricated the substrates; S.H.L. grew and characterized the crystal under the supervision of Z.Q.M.; all authors viewed and commented on the manuscript; V.V.D. supervised the whole project.

### Notes

The authors declare no competing financial interest.

## ACKNOWLEDGMENTS

This material is based upon work supported by the National Science Foundation the Quantum Leap Big Idea under Grant No. 1936383. K.L.W. acknowledges support from the U.S. Army Research Office MURI program under Grants No. W911NF-20-2-0166 and No. W911NF-16-1-0472. The financial support for sample preparation was provided by the National Science Foundation through the Penn State 2D Crystal Consortium-Materials Innovation Platform (2DCC-MIP) under NSF cooperative agreement DMR-2039351.

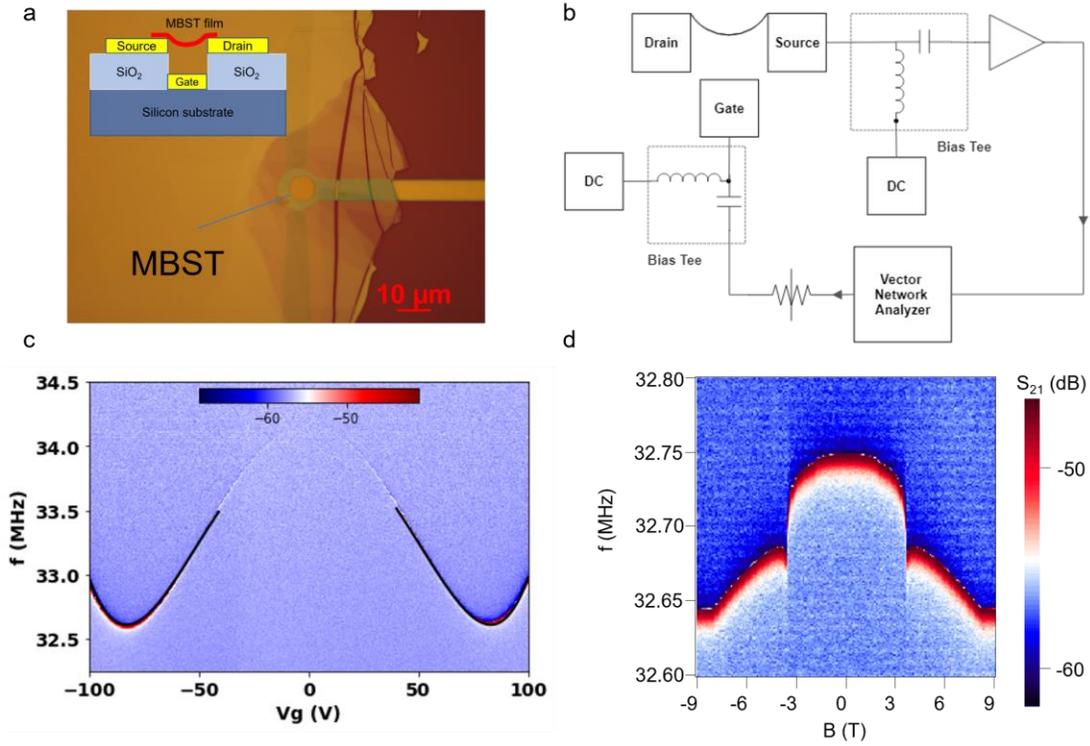

**Figure 1.** MBST NEM resonator. (a) Optical image of an MBST resonator device, insert plot is the side view of device geometry, the MBST film was suspended over the back gate. (b) Schematic of RF measurement circuit setup. The radio frequency (RF) excitation from the VNA output and DC voltage from Keithley are combined with a bias tee and then applied to the gate. A DC bias voltage can be applied to the source through the second bias tee. An amplifier is used before the RF signal feeds back into the VNA input to reduce the noise. (c) Gate dependence of resonance frequency at 1.5 K. The measured signal is $S_{21}$ (dB), and the amplitude of $S_{21}$ is shown in the color scale. The solid black line is the fitted curve from the continuum mechanical model with initial parameters of mass density mass ~7.169 kg/m$^3$, Young's modulus E = 84.374 GPa, and Poisson's ratio = 0.212. (d) Transmission signal $S_{21}$ as a function of driving frequency in the out-of-plane magnetic, the spin-flop transition is detected in the field dependence of resonance frequency shift.

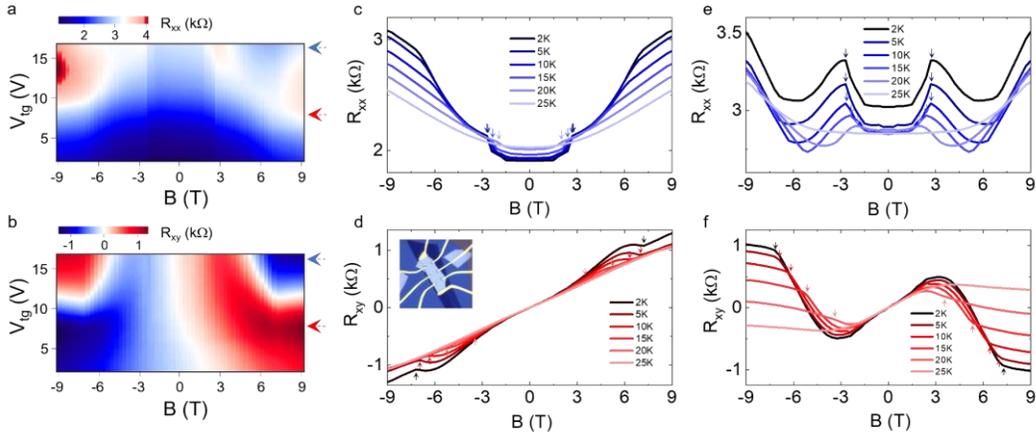

**Figure 2**. Color maps of (a) $R_{xx}$ and (b) $R_{xy}$ as functions of magnetic field and gate voltage for a 30 nm thick MBST flake. The data were taken at a temperature of 2 K. The blue and red horizontal arrows point to the topgate voltages where the line cuts are taken. Plots of (c) $R_{xx}$ and (d) $R_{xy}$ as a function magnetic field taken at different temperatures with the topgate voltage fixed of +7.8 V. Inset in (d) is the optical image of the MBST device. Plots of (e) $R_{xx}$ and (f) $R_{xy}$ as a function magnetic field taken at different temperatures with the topgate voltage fixed of +16.5 V. The $R_{xx}$ and $R_{xy}$ line profiles are symmetrized and anti-symmetrized, respectively. Vertical arrows in (c,e) and (d,f) point to the $H_{c1}$ and $H_{c2}$, respectively, at different temperatures.

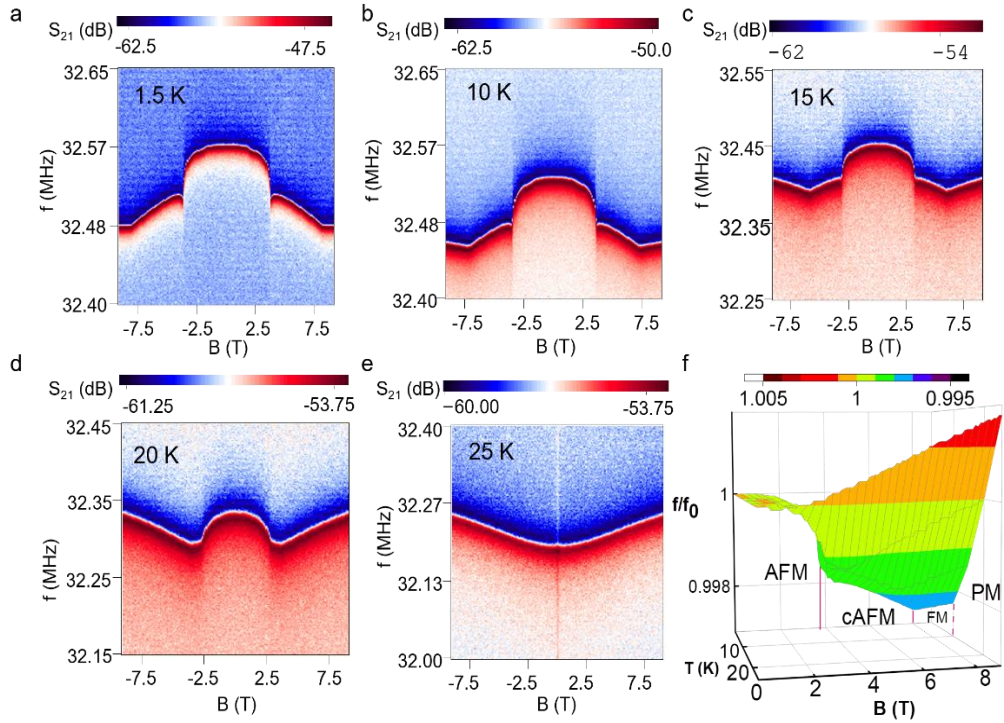

**Figure 3.** Field dependence of resonance frequency at different temperatures, at gate = 85 V. (a) - (e) are the color map of RF signal vs. driving frequency under an out-of-plane magnetic field for 1.5 K, 10 K, 15 K, 20 K, 25 K respectively. (f) The 3D plot of the resonance frequency is extracted from (a) to (e) as a function of magnetic field and temperature. Projections on the different planes depict the magnetic phase transition from AFM to canted AFM, and to FM transition in the low-temperature limit, and FM to PM phase transition in the high magnetic field limit (9 T).

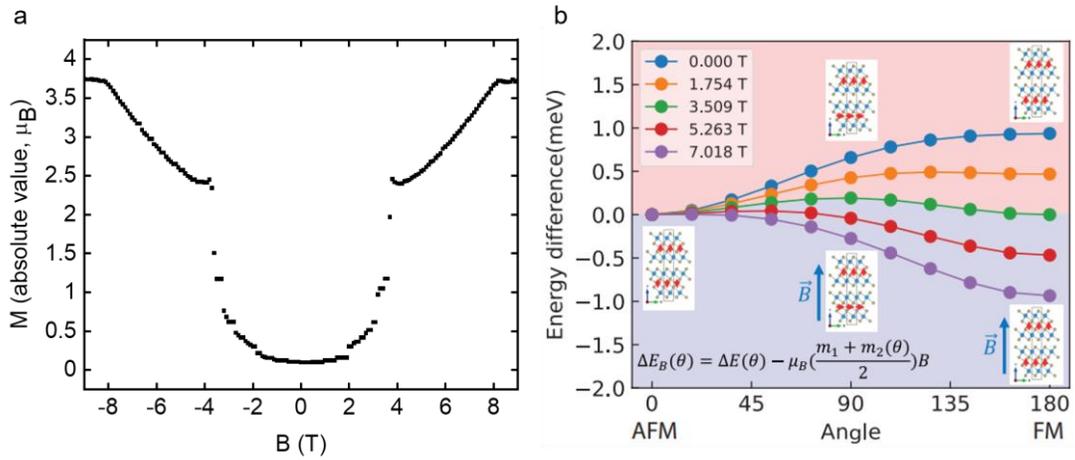

**Figure 4**. Effect of external magnetic field inducing the magnetic phase transition of MBST. (a) Magnetization (absolute value) per Mn atom as a function of out-of-plane magnetic field. (b) Relative energy of magnetic configuration as a function of spin angle in the bottom layer under the external magnetic field. (Zero energy is set to AFM ground state without external field). The effect of the external magnetic field is represented as colored solid lines. Red and blue shaded areas indicate where the AFM and FM state are more stable, respectively.